\documentclass[a4,aps,prl,twocolumn,amsmath,amssymb,superscriptaddress, showpacs, twoside,epsfig]{revtex4}

\usepackage{graphicx}
\usepackage{epstopdf} 
\newcommand{\be}[1]{ \begin{eqnarray} \mbox{$\label{#1}$} }

\newcommand{\ee}{\end{eqnarray}}
\newcommand{\pref}[1]{(\ref{#1})}

\newcounter{mycount}

\newcommand\ie {{\it i.e. }}
\newcommand\eg {{\it e.g. }}

\newcommand{\av}[1]{\langle #1\rangle}

\begin{document}

\title{Conformal Field Theory of Composite Fermions}
\author{T.H. Hansson} 
\affiliation {Department of Physics, Stockholm University
AlbaNova University Center,
SE - 106 91 Stockholm, Sweden} 
\author{C.-C. Chang}
\author{J.K. Jain}
\affiliation{Physics Department, 104 Davey Lab, The Pennsylvania State University, 
University Park, Pennsylvania 16802}
\author{ S. Viefers}
\affiliation{ Department of Physics, University of Oslo, P.O. Box 1048 Blindern, 0316 Oslo, Norway}

\date{\today}

\begin{abstract} 
We show that the quantum Hall wave functions  for the ground states in the Jain series can be 
{\em exactly} expressed in terms of  correlation functions of local vertex operators, 
$V_n$, corresponding to composite fermions 
in the $n^{\rm th}$ composite-fermion (CF) Landau level.  This allows for the 
powerful mathematics of conformal field theory to be applied to the successful CF phenomenology.
Quasiparticle  and quasihole states are expressed as correlators of anyonic operators with fractional (local) charge, allowing a simple algebraic understanding of their topological properties that are not manifest in the CF wave functions.

\end{abstract}

\maketitle

\newcommand{\CFT}{conformal field theory }
\newcommand\jas[3]{(z_{#1} - z_{#2})^{#3}}
\newcommand\sjas[3]{z_{#1#2}^{#3}}
\newcommand\prs[2] {\prod_{#1}\!^{(#2)}}
\newcommand\pr[3] {\prod_{#1<#2}\!^{(#3)}}
\newcommand\prjas[4] {\pr #1 #2 {#3}  \jas #1 #2 #4}
\newcommand\vmea[1] {e^{i\sqrt{m} \varphi (z_#1)}} 
\newcommand\vtmea[1] {\partial_{z_#1} e^{i(\sqrt{m}-\frac 1 {\sqrt m}) \varphi (z)}}
\newcommand\vme {e^{i\sqrt{m} \varphi_1 (z)}} 
\newcommand\vtme {\partial e^{i(\sqrt{m}-\frac 1 {\sqrt m}) \varphi (z)}}
\newcommand\vm[1] {V_1(z_{#1}) }
\newcommand\bvm[1] {V_0(\overline z_{#1}) }
\newcommand\vtm[1] { P_{\frac 1 m}(z_{#1}) }
\newcommand\bhvtm[1] { \hat P_{\frac 1 m}(\overline z_{#1}) }
\newcommand\hvtm[1] { \hat P_{\frac 1 m}(z_{#1}) }
\newcommand\bvtm[1] { P_{\frac 1 m}(\overline z_{#1}) }
\newcommand\holee[1]{e^{ \frac i {\sqrt m} \varphi (\eta_{#1} ) } }
\newcommand\hole[1] {H_{1/m}(\eta_#1)}
\newcommand\tphie{\tilde\varphi (z)}
\newcommand\tphi[1]{\tilde\varphi (z_{#1})}
\newcommand\eb{\overline\eta}
\newcommand\beb{\overline N}
\newcommand\dt[1]  {\tilde{\tilde {#1}}}

 In an intriguing line of development, many 
of the fractional quantum Hall effect (FQHE) wave functions, both ground states and 
the associated quasihole states, have 
been expressed as certain correlators in two dimensional conformal field theories (CFTs).  
Among these are the Laughlin states\cite{Laughlin83} at Landau level 
fillings $\nu=1/m$ ($m$ being an odd integer) and the Moore-Read Pfaffian wave function\cite{MR}
at $\nu=1/2$.  While there is no fundamental understanding of why the quantum mechanical 
wave functions of a 2D electron gas in the lowest Landau level (LLL)   
should bear any relation to correlation functions of vertex operators in a 2D Euclidian CFT, many arguments for such a connection were given in the pioneering 
paper by Moore and Read\cite{MR} and discussed and extended in several subsequent works\cite{froh}.

The charged excitations in the FQH systems in general
have a fractional charge\cite{Laughlin83} and exhibit
fractional statistics\cite{Halperin84} - they are anyons\cite{Leinaas}.  
 These are a manifestation of 
subtle long range quantum correlations that characterize the FQH states, 
which are said to be topologically ordered. 
When several such excitations form composites, their charges and statistical 
properties can be deduced from their constituents using {\em local} rules. 
The fractional excitations as well as these  local rules of ``fusion" 
and the related ``braiding" (\ie statistics) properties of the FQH states find their 
counterpart in CFTs. The excitations are represented by operators, 
the fusion corresponds to their
operator product expansion (OPE), and the braiding rules follow 
from certain algebraic properties
(monodromies) of their correlation functions (or conformal blocks).

There are, however,  important limitations to the CFT approach.  
A connection between CFT and quasiparticle (as opposed to quasihole) 
excitations at $\nu=1/m$ has so far been lacking\cite{froh}.  
Furthermore, in spite of interesting progress\cite{flohr}, 
no CFT expressions have so far  been established  for  the experimentally prominent 
FQH states in the Jain series at filling fractions $\nu=n/(2np+1)$, and their
quasiparticle or quasihole excitations.

The composite fermion (CF) formalism  provides for 
a unified description of the Jain ground states and their quasiparticle and quasihole excitations. 
A composite fermion is an electron bound to $2p$ units of vorticity, and the ground states are $n$ filled 
CF or "effective", Landau levels. The phenomenology based on composite fermions has been successful, 
both in comparison with experiments and with various exact numerical studies\cite{reviews}.

CF wave functions for quasiparticles and quasiholes are obtained by adding 
a composite fermion to, or removing it from, 
a CF Landau level.   Although phrased in a single particle language, 
this is a non-local operation in terms of electrons, and 
qualitatively different from the corresponding local\footnote{
Or, more precisely, {\em quasi}-local since  LLL  projection destroys locality on the scale of the magnetic length $\ell$. }
process in an integral QH state. Thus  
simplistic analogies can lead to qualitatively wrong conclusions.
While fractional charge and fractional braiding statistics can be derived
from the pertinent CF wave functions, these topological properties are 
not manifest.

In this Letter, we present an exact 
representation of the CF wave functions  in the Jain series as  correlators of vertex operators
in a CFT involving $n$ compactified boson fields\cite{hcjvl}.  Quasiparticle and quasihole excitations are described by operators that explicitly obey anyonic statistics and 
have fractional charge. 
This deepens and extends the CFT formulation of the 
FQHE, and relates the topological properties of the excitations 
to algebraic properties of vertex operators.  The very same CFTs  
are also the pertinent edge theories for the Jain states, giving a direct link between CF
states and the $K$-matrix approach to QH edges developed by Wen\cite{wen}.

We first recall how to construct the ground state and quasihole  wave functions at the Laughlin 
fractions $\nu =  1/m$. 
Following ref. \onlinecite{MR}, we define the vertex operators,
\be{vo}
V_{1}(z) &=& \vme \\
H_{\frac 1 m}(\eta) &=& e^{ \frac i {\sqrt m} \varphi_1 (\eta) } \, , \label{hole}
\ee
with  $z=x+iy$. The field
$\varphi_1$ is a free massless boson normalized as to have the (holomorphic) two point function
$
\av{ \varphi_1 (z) \varphi_1 (w) } = - \ln (z - w) 
$.
$\varphi_1$ is compactifed on a circle of radius $R^2 =  m$ so the (holomorphic) $U(1)$ charge density operator is
$
 J(z) =\frac i {\sqrt m} \partial \varphi_1 (z)   \, ,
$
with $\partial =\partial / \partial z$.
The corresponding $U(1)$ charge, ${\cal Q} = \frac 1 {\sqrt m} \frac 1 {2\pi } \oint dz\,  \partial \varphi_1 (z) $, 
which can be thought of as vorticity, 
 can be read directly from the commutators, 
$[{\cal Q}, V_{1}(z) ] = V_{1}(z) $ and $ [{\cal Q}, H_{1/m}(\eta) ] = 
\frac 1 m H_{1/m}(\eta)  $, and
equals 1 for  the electron and $1/m$ for the hole.
Following the basic idea of  Read\cite{read}, the total {\em excess electric charge} associated with a vertex operator, $Q_{el} = e(Q - \delta n)$,
gets contributions both from the electrons actually added, $\delta n$, and the change in vorticity, $Q$, corresponding to a depletion or contraction of the electron liquid.
 If the argument of the vertex operator is an electron coordinate, $z_i$, one electron is added, while no electron is added if the argument is a hole coordinate $\eta_i$. 

The $\nu = 1/m$ Laughlin ground state wave 
function, 
$
\Psi_L(\{z_k\}) \equiv \Psi_L  = \prod_{j<k}(z_j-z_k)^m\, 
$
can be written\footnote{
The proper definition of the average $\av{\dots}$ involves a neutralizing background charge\cite{MR}, not explicitly shown here. 
The operators are assumed to be radially ordered.  }
\be{La}
\Psi_L  &=&      \av{ \vm 1 \vm 2 \dots \vm {N-1} \vm N } \, .
\ee
 The wave function for a collection of Laughlin quasiholes is given by similar expressions containing insertions of the hole operator $H_{1/m}$.

The most natural guess\cite{MR} for a quasiparticle operator would be 
to change the sign in the exponent in the hole operator of  \pref{hole}. 
This, however, introduces singular terms $\sim \prod_i (z_i - \eta)^{-1}$ in the 
electronic wave function. Inspired by the CF wave functions, we instead define a 
quasiparticle operator, $ P_{1/m}(z)$, by  modifying one of the electron operators to have 
$U(1)$ charge  $(1- 1/m)$. The excess electric charge associated with such a modification is the difference between the charges of the operators $V_{1}$ and $P_{1/m}$ \ie    $\Delta Q_{el} =  e((1-1/m)-1) = -e/m$ appropriate for a Laughlin quasiparticle. The modified electron operator is given by 
\be{qpo}
 P_{\frac 1 m }(z) = \vtme \, . 
\ee
The  wave function for a single quasiparticle in angular momentum $l$  takes the form 
(from now we will suppress the gaussian factor $\exp-\{ \sum_i |z_i|^2 /(4\ell^2)  \}$, where $\ell^2 = \hbar/eB$.) , 
\be{1qp}
\Psi_{1qp}^{(l)}  =  {\cal A}\{  z_1^l   e^{- |z_1|^2 /4m\ell^2}  \av{ \vtm 1 \prod_{i=2}^N\vm i     } \} \nonumber   \\ 
= \sum_i (-1)^i z_i^l \pr j k {i} (z_{j} - z_k)^m 
\partial_i  \prs l i (z_{l} - z_i)^{m-1} \
\ee
where ${\cal A}$ denotes anti-symmetrizaion of the coordinates, and 
the superscript of $\prod$ indicates indices omitted in the 
product (\eg  
$
\prs i j = \prod_{\stackrel {i=1} {i\ne j}}^N
$  and 
$
\pr i j {kl}  = \prod_{\stackrel {i< j} {i,j\ne k,l}}^N  $).
The last product in \pref{1qp} comes from contracting the operators $P_{1/m}$ and $V_1$;
since $(m-1)$ is even, the derivative is important to ensure 
that antisymmetrization does not annihilate the state. Put differently, 
contracting the electron liquid around the point $z_i$ amounts to having a leading short distance  part in the 
wave function $\sim (z_{l} - z_i)^{m-2}$, as seen from the OPE  $P_{1/m}(z) V_{0}(w) \sim (z-w)^{m-2} $, where the power changes from $(m-1)$ to $(m-2)$ because of the derivative.

The wave function in \pref{1qp} is {\em identical} to the corresponding CF result 
(Eq. 5 of ref. \onlinecite{jain3}) 
and thus has a good variational energy and the correct fractional charge. 
A localized quasiparticle state
can be constructed as a coherent superposition of the angular momentum eigenstates 
of   \pref{1qp}.  Note that our construction is entirely within the LLL -- no 
projection is needed\cite{hax}. 
The operator $ P_{1/m}(z) $ is not fermionic, as can be seen from the
OPE  $P_{1/m}(z) P_{1/m}(w) \sim (z-w)^{(m-4+\frac 1 m)} $.
The precise connection to composite fermions is clarified below. 

The quasiparticle wave function of 
\pref{1qp} has a different character than those written 
earlier for the ground and the quasihole states, in that it is a sum over correlators and that it 
involves a prefactor.  The 
factor $z^l$ gives the correct angular momentum, and the exponential factor is chosen 
to give the correct LLL  electronic wave function. It is suggestive that these prefactors precisely constitute the LLL wave function $f_1 = z_i^l   e^{- |z_i|^2 /4m\ell^2}$ for a charge $e/m$ particle in the LLL. Although we have no fundamental derivation of 
this, we find below a similar interpretation  in the case of several quasiparticles, 
where their anyonic nature is also manifest.

A natural generalization to two quasiparticles is
\be{2qp}
\Psi_{2qp} ={\cal A} \{  f_2(z_1,z_2)  
   \langle  \vtm 1  \vtm 2  \prod_{i=3}^N\vm i     \rangle  \} \ .       
\ee
The correlator gives a factor $\partial_i\partial_j\jas i j {m-2+1/m}$, so to get an analytic electron wave function with the limiting behavior $\sim \jas i j {m-1+l}$ with $l\ge 1$ and odd, we should take 
$f_2(z_i,z_j) = g(Z_{ij}) \jas i j {1+l-\frac 1 m} e^{- (|z_i|^2+|z_j|^2) /4m\ell^2}$,
where $Z_{ij} = (z_i + z_j)/2$. This is precisely the expected wave function for two fractionally charged anyons in the LLL.
Evaluating the correlator  gives the explicit CFT wave function for 
two quasiparticles in relative angular momentum $l$ and center of mass angular momentum $L$,
\be{2qpe}
\Psi_{2qp} = \sum_{i<j} (-1)^{i+j} Z_{ij}^L  \sjas i j {1+l-\frac 1 m}  \partial_{i} \partial_{j} \sjas i j {m-2+ \frac 1 m}  \nonumber \\
 \prs k  {j}  \sjas k i {m-1}   \prs p  {i} \sjas p j {m-1}  \prs {q< r} {ij}  \sjas q r m
\ee
where $z_{ij} = z_i - z_j$ and the derivatives act on the whole expression to their right. 
The first nontrivial test of our construction is whether \pref{2qp}  produces accurate wave functions.
To investigate that, we compare this CFT wave 
function with the standard CF wave function\cite{jain3}.
The latter is similar in structure, but not identical;  it can be obtained from  \pref{2qpe} 
by replacing $\sjas i j {1+l-\frac{1}{m}}\partial_{i} \partial_{j} \sjas i j {m-2+ \frac 1 m} $ with 
$\sjas i j {l}\partial_{i} \partial_{j}  \sjas i j {m-1}$.  The comparisons, using the 
Metropolis Monte Carlo algorithm, for a quasiparticle pair with 
$L=0$ and $l=1$ are summarized in Table I and Figure 1\footnote{
 Note that the angular momentum 
 differs from that of two Laughlin quasiparticles,
$\Psi_{L,2qp}\sim \prod_j \partial_j^2\Psi_L$ \cite{jain3}. }.
The almost perfect agreement between the two wave functions is surprising.

\begin{table}
\begin{center}
\begin{tabular}{rccc}\hline\hline
$N$   &   $E_{\rm CFT}$      &    $E_{\rm CF}$        &    $\langle\Psi_{2qp}|\Psi_{2qp}^{\rm CF}\rangle$\\\hline
  10    &  7.76619(62)    &   7.76600(62)  & 0.9999301(2)   \\
  20    & 24.1403(19)     &  24.1402(19)    &  0.9999274(4)  \\
  30    & 46.3258(18)     &  46.3257(18)    & 0.9999274(3)   \\
  40    & 73.2339(17)     &  73.2339(17)    & 0.9999266(2)   \\
\hline\hline
\end{tabular}
\end{center}
\caption{The expectation values of the Coulomb Hamiltonian with respect to the 
two-quasiparticle CFT wave functions in \pref{2qpe} ($E_{\rm CFT}$), with 
$L=0$ and $l=1$,  
and the corresponding CF wave function in Eq. 14 of Ref. \onlinecite{jain3} ($E_{\rm CF}$).
The energies are quoted in units of $e^2/\ell$, $\ell=\sqrt{\hbar c/eB}$.
The last column gives the overlap between the two wave 
functions.  $N$ is the number of particles.}
\end{table}

\begin{figure}
\includegraphics[scale=0.57,angle=-90]{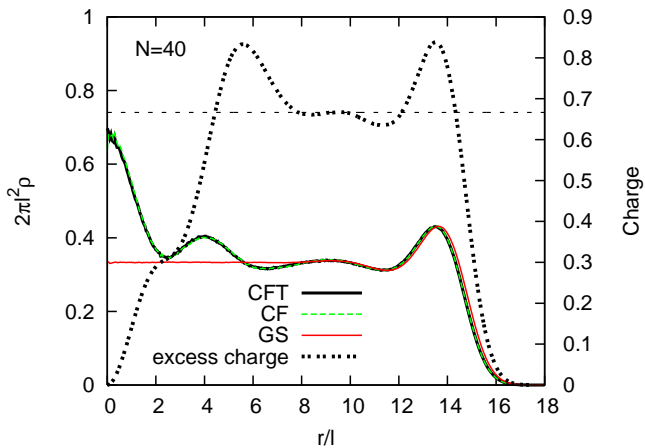}
\caption{The density profile for the CFT state in \pref{2qpe} which represents two 
quasiparticles at $\nu=1/3$, along with the standard CF state.  
The total number of electrons is $N=40$.  The excess 
charge (above the 1/3 ground state labeled by GS) inside the disk of radius $r$ 
has a clear 
plateau at 2/3 outside the region containing the two quasiparticles, and also away 
from the edge. The radius $r$ is shown in units of the magnetic length, and the density 
$\rho$ in units of the $\nu=1$ density $(2\pi l^2)^{-1}$.
\label{hansN40}}
\end{figure}

The properties of the operator $P_{1/m}$, as well as the 
analytic form of the wave function in  \pref{2qp}, strongly suggest that the 
CFT quasiparticle has good statistics properties, which is also all but guaranteed by the 
near identity to  the corresponding CF wave function. 
However, an analytical derivation of the quasiparticle statistics 
is both of intrinsic interest and a demonstration of the power of the CFT approach. 
In the case of quasiholes, charge and statistics can be extracted from Berry phases evaluated 
using the Laughlin plasma analogy, or a generalization thereof based on properties of CFT correlators, proposed by Nayak and Wilczek\cite{nw}. 
For quasiparticles it is less straightforward, since the
electronic wave functions obtained from Eqs. \pref{1qp} and \pref{2qp} 
involve sums over correlators. 
Nevertheless, the CFT expressions are simple enough that the Berry phases can be evaluated analytically
and the results for the fractional charge and statistics of quasiparticles  are as expected.\footnote
{This relies on a 
RPA,  in which the off-diagonal terms of the density matrix are neglected, 
More details and the justification are left for a longer 
paper \cite{hcjvl}. }

We now generalize to 
$M$ quasiparticles at $\nu=1/m$.  We consider a maximum density circular droplet 
obtained by putting all the quasiparticle pairs in their lowest allowed relative angular 
momentum, and with zero angular momentum for the center of mass. For simplicity we 
now specialize to $m = 3$. 
The resulting wave function
\be{nqp}
\Psi_{Mqp}   =  \sum_{i_1<i_2 \dots  < i_M} (-1)^{\sum_k i_k}  
\prod_{k<l}^M  \jas {i_k} {i_l}  {\frac 5 3}   \nonumber \\ 
e^{- \sum_i^M|z_i|^2 /4m\ell^2}
 \langle 
 P_{\frac 1 3 } (z_{i_1})   \dots P_{\frac 1 3} (z_{i_M})\prod_{j\in\!\!\!  | \, \{i_1\dots i_M\} }^N V_1(z_ j)     \rangle \, 
\ee
again differs from the CF wave function only in the ordering of the derivatives and the 
Jastrow factors, and can be expected to be accurate.

This suggests an exact representation of the CF wave functions as CFT correlators
by introducing several bosonic fields.
The factor $\prod_{k<l}^M(z_{i_k}-z_{i_l})^{5/3}\exp(- \sum_i^M|z_i|^2 /4m\ell^2)$ in \pref{nqp}, which has the interpretation of  
an $M$-quasiparticle wave function $f(z_i)$ of a maximum density droplet of anyons in the LLL, 
can be expressed as a correlator of vertex operators of a second free bosonic field $\varphi_2$. 
Define 
\be{vtil}
V_{2}(z)  = \partial  e^{i\frac 2 {\sqrt 3} \varphi_1(z)}  e^{i\sqrt{\frac 5 3}\varphi_2(z)}  \, ,
\ee
and consider the correlator
\be{nqpc}
\Psi_{Mqp}  &=& \sum_{i_1<i_2 \dots < i_M} (-1)^{\sum_k^M i_k} \nonumber \\  
 \langle 
V_{2} (z_{i_1})   &\dots&  V_{2} (z_{i_M})\prod_{j\in\!\!\!  | \, \{i_1\dots i_M\} } V_{1}(z_ j)     \rangle 
\ee
of $M$ $V_{2}$:s and $(N-M)$ $V_{1}$:s. 
If we had chosen $V_{2}(z)  = e^{i\sqrt{\frac 5 3}\varphi_2(z)}\partial   e^{i\frac 2 {\sqrt 3} \varphi_1(z)}$,   \pref{nqpc} 
would have reproduced  \pref{nqp}.  With the choice in 
 \pref{vtil}, we obtain instead:
\be{nqpc2}
\Psi_{Mqp}  && =\sum_{i_1<i_2 \dots < i_M} (-1)^{\sum_m^M i_m}\partial_{i_1}\cdots 
\partial_{i_M} \\
&& \prod_{i_l<i_m}(z_{i_l}-z_{i_m})\prod_{\bar i_l<\bar i_m}(z_{\bar i_l}-z_{\bar i_m})
\prod_{j<k}(z_j-z_k)^2\;. \nonumber
\ee
Here, the indices $j$ and $k$ run over all particles, $\{i_m\}$ represent a subset of $M$ indices, and $\{\bar i_m\}$ represent the remaining $N-M$ indices.  
While one can write general $M$-quasiparticle wave functions similar to the two particle case \pref{2qp}, 
it is only the maximum density droplet \pref{nqp} that allows for a simple expression in 
terms of conformal blocks.\footnote{
 This is related to the necessity to introduce a neutralizing background charge 
in order to get non vanishing correlators. For small deviations from the $\nu = 1/3$ state one can 
artificially introduce compensating charges related to the field $\varphi_2$, but there is no natural way to do this. }  
Using $[\varphi_1(z) , \varphi_2(w)]=0$ and $\av{ \varphi_i (z) ,  \varphi_i (w)} = - \ln (z - w) $ one can show 
that the $V_2$'s, just as the $V_1$'s but different from the  $P$'s, are fermionic operators, 
satisfying $V_2(z_i)V_2(z_j)+V_2(z_j)V_2(z_i) = 0$, and also that $V_{1}$ and $V_2$ commute. 
If we want to interpret $V_2$ as a composite {\em electron} operator, it should have the same charge as $V_1$. This is ensured if we redefine the charge density operator as 
$
J(z) =\frac i {\sqrt 3} \partial \varphi_1 (z)  +   \frac i {\sqrt{15}} \partial \varphi_2(z)
$, consistent with the field $\varphi_2$ being compactified with radius $R^2 = 15$. 

Taking an equal number of $V_{1}$'s and $V_2$'s, \ie  $N=2M$, charge neutrality implies $\nu = 1/ 3 + 1/ {15} = 2/5$. 
To see that  \pref{nqpc2} in this case {\em exactly} reproduces  the CF  wave function  for the $\nu = 2/5 $ state, we note that the latter is given by 
\be{nqpc3}
\Psi_{Mqp}  ={\cal P}_{\rm LLL} 
\left|  
\begin{array}{cccc}
\bar z_1  &  .& .& \bar z_N  \\
\bar z_1  z_1 &  .& .& \bar z_N  z_N \\
. & .&  .& . \\
\bar z_1 z_1^{M-1} & .& .& \bar z_N z_N^{n-1} \\
1 &  .& .& 1 \\
z_1   & .& .& z_N \\
. & .& .& . \\
z_1^{M-1} &  .& .& z_N^{M-1} \\
\end{array} \right |  \prod_{j<k}(z_j-z_k)^2\;.  \nonumber
\ee
Using the Laplace expansion by the first $M$ rows of the determinant, and then carrying 
out the lowest LL projection (indicated by the symbol ${\cal P}_{\rm LLL}$) by moving all 
of the $\bar z$'s to the left and making the replacement $\bar z_j\rightarrow 2\partial_j$ 
reproduces  \pref{nqpc2}, apart from a normalization factor. This proof trivially generalizes to  $m = 2p +1$ yielding the $\nu = 2/(4p +1)$ states.

To create quasiholes in the 2/5 state the operator $H_{1/3}(\eta)$ given in \pref{hole} 
is no longer appropriate since it does not give holomorphic electron wave functions. 
That is accomplished by using either of 
\be{hmod}
H_{2/5} &=&   
   e^{ \frac i {\sqrt 3} \varphi_1 (\eta) + \frac i {\sqrt {15} } \varphi_2(\eta) }        \\
H_{1/5} &=&    
=    e^{ \frac i {\sqrt 3} \varphi_1 (\eta) -\frac {2i} {\sqrt {15} } \varphi_2(\eta) }  \, ,  \nonumber
\ee
with charges 2/5 and 1/5 respectively. 
The explicit electron wave functions are obtained by inserting the operators \pref{hmod} in the correlator \pref{nqpc}.

By introducing more bosonic fields, the above construction can be generalized to wave 
functions for compact states of CFs involving still higher CF Landau levels.  
For example, the vertex operator for placing composite fermions in the third  
CF-LL is given by\cite{hcjvl}
\be{vtiltil}
V_{3}(z)  = \partial^2   e^{i\frac 2 {\sqrt 3} \varphi_1(z)}  e^{i\frac{2}{\sqrt{15}}\varphi_2(z)} \ e^{i\sqrt{\frac 7 5}\varphi_3(z)} \, .
\ee
A general state in the Jain series $\nu=n/(2np+1)$ requires  $n$ different fermionic operators $V_{s}$, $s=1\dots n$, 
which are constructed  from $n$ free, compact, boson fields, $\varphi_s$, and $n$ derivatives. 
Again the pertinent antisymmetrized CFT correlator exactly reproduces the corresponding CF wave functions. 
This also provides for a natural connection between the CF theory and Wen's general description of 
topological fluids based on Chern-Simons actions, and hence candidate edge theories for the CF states.  
We have shown\cite{hcjvl} that for the $\nu = 2/5$ and $\nu = 3/7$  
states the effective theory based on the fields $\varphi_i$,  $i = 1,2,3$, exactly reproduces Wen's results\cite{wen}. 
The form \pref{hmod}, for the quasihole operators is essential for this correspondence.

We are grateful to J.M. Leinaas and A. Karlhede for insightful discussions and comments. 
We thank Prof. B. Janko for hospitality at the ITS, a joint institute of ANL and the University of
Notre Dame, funded through DOE contract W-31-109-ENG-38 and Notre Dame
Office of Research. 
 This work was supported in part by VR, by Nordforsk, and by NSF under grant no. DMR-0240458.

\end{document}